\documentclass[conference]{IEEEtran}

\IEEEoverridecommandlockouts
\usepackage{verbatim}
\usepackage{longtable}
\DeclareMathAlphabet{\mathpzc}{OT1}{pzc}{m}{it}
\usepackage{amsmath}
\usepackage{fancyhdr}
\usepackage{amsfonts}
\usepackage{amssymb}
\usepackage{multirow}
\usepackage{amsmath,amssymb,amsfonts}
\usepackage[ruled,vlined,linesnumbered]{algorithm2e}
\usepackage[
top    = 0.7in,
bottom = 0.95in,
left   = 0.65in,
right  = 0.57in]{geometry}
\usepackage{graphicx}
\usepackage{textcomp}
\usepackage{xcolor}
\usepackage{subcaption}
\usepackage{algpseudocode}
\usepackage{enumerate}

\begin{document}

\title{Quarks: A Secure and Decentralized Blockchain-Based Messaging Network}


\author{\IEEEauthorblockN{Mirza K. B. Shuhan$^{1}$, Tariqul Islam$^{2}$, Enam A. Shuvo$^{3}$, Faisal H. Bappy$^{4}$, Kamrul Hasan$^{5}$, and Carlos Caicedo $^{6}$}
\IEEEauthorblockA{
$^{1}$ Software Research \& Engineering, bKash Limited, Dhaka, Bangladesh \\
$ ^{2, 4, 6}$ School of Information Studies (iSchool), Syracuse University, Syracuse, NY, USA\\
$^{3}$ School of Computing, Asia Pacific University of Technology and Innovation, Kuala Lampur, Malaysia\\
$^{5}$ Tennessee State University, Nashville, TN, USA\\
Email: $\lbrace$\textit{shuhan.mirza@gmail.com, mtislam@syr.edu, enam.ahmed.shuvo@gmail.com, fbappy@syr.edu,} \\ \textit{mhasan1@tnstate.edu, ccaicedo@syr.edu}$\rbrace$}
}

\maketitle

\begin{abstract}

Over the past two decades, the popularity of messaging systems has increased both in enterprise and consumer level. Many of these systems used secure protocols like end-to-end encryption to ensure strong security features such as ``future secrecy" for one-to-one communication. However, the majority of them rely on centralized servers owned by big IT companies, which allows them to use their users' personal data. Also it allows the government to track and regulate their citizens' activities, which poses significant threats to ``digital freedom". Also, these systems have failed to achieve security attributes like confidentiality, integrity, privacy, and future secrecy for group communications. In this paper, we present a novel blockchain-based secure messaging system named \texttt{Quarks} that overcomes the security pitfalls of the existing systems and eliminates the centralized control. We have analyzed our design of the system with security models and definitions from existing literature to demonstrate the system's reliability and usability. We have developed a Proof of Concept (PoC) of the \texttt{Quarks} system leveraging Distributed Ledger Technology (DLT), and conducted load testing on that. We noticed that our PoC system achieves all the desired attributes that are prevalent in a traditional centralized messaging scheme despite the limited capacity of the development and testing environment. Therefore, this assures us the applicability of such systems in near future if scaled up properly.

\end{abstract}

\begin{IEEEkeywords}
instant messaging, blockchain, group communication, system security.
\end{IEEEkeywords}

\thispagestyle{fancy}
\lhead{This work has been accepted at the 10th IEEE International Conference on Cyber Security and Cloud Computing (CSCloud 2023)}
\cfoot{}

\section{Introduction}


The popularity of Messaging Systems has been increasing for its prompt response time, convenient user experience, and ease of multi-tasking in both informal and formal communication and collaboration. Despite it's uprising popularity, there is also some concerns about the resiliency of these services and users' control over data. Because all popular messaging systems leverage central servers to route text messages \cite{instantMessagingTraffic}. This gives more control to big tech companies and government to provision the user activities.

The Internet's provision of digital freedom is not a new thing. It started with the use of metadata by big tech companies for profit-making opportunities. Centralized systems provide more control over user personal data, which can be sold to third parties or used for targeted marketing. Moreover, centralized systems are frequently monitored by governments in order to keep track of their citizens' activities, and some countries impose regulatory laws that potentially jeopardize user privacy. Telegram, for example, was banned in Russia in April 2018 because it refused to provide the Russian Federal Security Service (FSB) with access to encrypted messages, as required by anti-terrorism laws \cite{wijermars2021selling}. Centralized architectures are more likely to pose the biggest threats to privacy and freedom of speech, including single point of failure, data leakage, and control over private conversations \cite{musiani2022concealing}. Therefore, the current Web 2 foundation cannot guarantee privacy or freedom of expression.


To tackle mass surveillance of conversations by government agencies and large corporations, all major messaging applications are integrating end-to-end encryption to their protocols, such as Signal \cite{signalE2E}, WhatsApp \cite{whatsappE2E}, Threema \cite{threemaE2E}, Google Allo \cite{alloE2E}, and Facebook Messenger \cite{fbE2E}. End-to-end encryption has been proven \cite{cohn2020formalE2E} to be an outstanding way to implement secure messaging protocol that ensures additional security properties like \textit{future secrecy} \cite{signalCryptographicRacheting}. However, we have seen these messaging systems are vulnerable to malicious attacks due to the improper implementations and imbalance between “usability first” or “privacy first”. These attacks include server-based attacks, such as vulnerability of Apple's iMessage  \cite{garman2016dancingImessage} and Signal Protocol \cite{kobeissi2017automatedSecMsgProto}. Moreover, researchers have been able to decipher the message database at end-users' devices of major messaging apps like WhatsApp, WeChat and Viber \cite{rathi2018forensicIM,gudipaty2015whatsappForensic, grover2013androidForensic,wu2017weChatforensic}. 

Conventional models of data security rely on creating harder and harder walls-- adding multiple factors of authentication to ensure access control and emphasizing stronger encryption. With Blockchain, there exists the potential to scatter the stack, rendering the cost of any one breach or combination of breaches much lower. Combined with strong encryption methods and zero knowledge proofs, Blockchain-based messaging systems can be a much more secure method of storing, accessing, and transmitting data; enhancing the ability of data managers to protect critical information.

\textbf{Research Questions.} The following research questions ($RQ$s) intrigued us to investigate existing messaging systems and conduct this research. $RQ_1$: Can blockchain-based messaging systems be utilized to ensure digital freedom and eliminate control over user data? $RQ_2$: Is it possible to build a blockchain-based messaging system that includes all the features of a traditional messaging system? and $RQ_3$: Can blockchain-based (decentralized) messaging systems overcome the security flaws of existing centralized messaging models?



In this paper, we try to address the above-mentioned research questions by proposing \texttt{Quarks}, a novel application of Blockchain in decentralized messaging system which has not been explored before. \texttt{Quarks} eliminates central control over data and ensures true decentralization, security, privacy, and trust. The following are the major contributions of the paper.

\begin{itemize}

\item We developed a PoC of \texttt{Quarks} system that ensures ``Message Integrity" and ``Trust in Federation" using Distributed Ledger Technology (DLT).

\item \texttt{Quarks} ensures digital freedom, future secrecy, and guarantees no single point of failure. 

\item We conducted performance and security analysis, which demonstrate that our PoC meets all intended features of a truly decentralized messaging system.




\end{itemize}

The rest of the paper is organized as follows. In Section \ref{sec:related_work}, we discuss some related work on the existing messaging systems. The architecture of the \texttt{Quarks} is presented in Section \ref{sec:architecture}. In Section \ref{sec:protocol}, we describe the protocol flow of our \texttt{Quarks} system. We present implementation, system performance, and informal security analysis of \texttt{Quarks} in Section \ref{sec:implementation}. Finally, we conclude the paper in Section \ref{sec:conclusion}.

\section{Related Works}
\label{sec:related_work}

Sarıtekin et al. \cite{saritekin2018blockchain} and Mirzaei et al. \cite{mirzaei2022simorgh} propose  communication applications named `CrypTouch' and `Simorgh' respectively, where both the schemes used IPFS (InterPlanetary File System) to store the messages off-chain. However, details about their network structure and protocol flow was not presented in the paper. Also, IPFS is suitable only for storing large files. In case of messaging systems, storing the messages in IPFS will add extra overhead as most of the messages are smaller in size. And it will require extra resources for each nodes to store and maintain the whole messaging system. 

A chat application using Ethereum's Whisper protocol is proposed by Abdulaziz et al. \cite{abdulaziz2018decentralized}. This protocol is designed to ``communicate darkness" (i.e., the content of the message is inaccessible to those who intercepts the messages, and that communicating nodes cannot be easily identified) at high cost. It is a off-chain protocol in which every message is sent to every node, and every node tries to decrypt the message. Consequently, the protocol has high traffic, high processor and memory usages. Hence, this is not an efficient protocol for chatting applications. Menegay et al. \cite{menegay2018secure} attempt to implement a communication application using `Steem', which is commonly referred as the ``social blockchain", designed to power blockchain-based blogging and social media platforms. However, ``Steem" is a public content platform which is not suitable for implementing communication application. A blockchain-based secure communication framework for community interaction is proposed by Sharma et al. \cite{sharma2021blockchain}. Their scheme manages identity of network's user through third party centralized service (Google) and all communication data are kept in centralized database. 

Currently, in most of the popular messaging systems, text messages are routed through central servers \cite{instantMessagingTraffic}. It makes the system prone to single-point of failure, although it provides the service-provider with fine-grained control over the system. Many companies, namely Signal \cite{signalE2E}, WhatsApp \cite{whatsappE2E}, Threema \cite{threemaE2E}, Google Allo \cite{alloE2E}, and Facebook Messenger \cite{fbE2E}, have servers that are maintained by the service providers and they have full access to the sensitive conversation data which raises privacy concerns. Furthermore, researchers have been able to intercept sensitive information from WhatsApp \cite{karpisek2015whatsappNetworkForensic, wijnberg2021interceptingWhatsapp, cents2020identifyWhatsAppMsg}, which has over two billion users worldwide. R{\"o}sler et al. \cite{rosler2018more} have analyzed the group communication of  Whatsapp, Threema, and Signal. Their analysis disclosed that the \textit{communication integrity} and the \textit{groups' closeness}
are not end-to-end protected. In addition, they proved that Signal protocol can not maintain strong security properties like \textit{future secrecy} in group communication.

\section{Background and System Architecture}
\label{sec:architecture}

In this section, we first review some preliminaries and then present our system architecture.

\subsection{Background}
\textbf{Blockchain.} Blockchain is a smartly engineered distributed system featuring an immutable ledger of transactions shared and validated by a number of distributed Peer-to-Peer (P2P) nodes \cite{chowdhury2019comparative}. The ledger is an ordered data structure consisting of many blocks chained together by cryptographic mechanisms. 

\textbf{Smart Contract (SC).} Smart Contracts are computer programs deployed on top of the respective blockchain \cite{ferdous2019search}. Being part of the ledger makes SC and their executions immutable and irreversible, a sought-after property having a wide range of applications in different domains. 

\textbf{Future Secrecy.} Future secrecy is a prime feature of key agreement protocols in messaging systems which prevents an adversary (i.e., who compromises the message 
keys of a target user) from decrypting any future messages in the conversation to some extent. This is achieved by a unique technique called ``ratcheting" in which session keys are updated with every message sent \cite{fSec}.

\subsection{\texttt{Quarks} Components}
The following four are the key components and participants of our proposed \texttt{Quarks} System. \textbf{i. User:} Users are the actors who use the system to communicate with their acquaintances; \textbf{ii. \texttt{Quarks} Channel:} A channel is a message thread between two or more users; \textbf{iii. \texttt{Quarks} Node:} A \texttt{Quarks} node independently hosts multiples users and their messages and \textbf{iv. \texttt{Quarks} Network:} A \texttt{Quarks} Network consists of multiple \texttt{Quarks} nodes where users from different nodes interact with each other.

\subsection{High Level View of \texttt{Quarks} System}


\begin{figure}[!h]
\centering
\includegraphics[width=2.8in, height=1.0in]{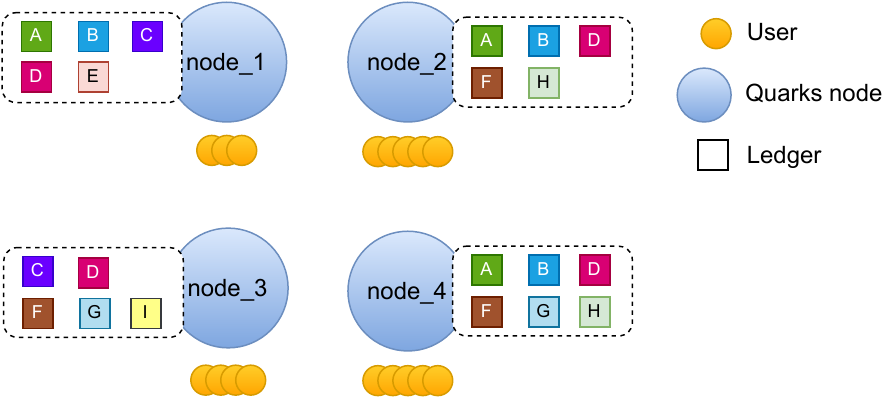}
\caption{High level view of \texttt{Quarks} System}
\label{fig:quark_arc}
\end{figure}

The high level semantics of the proposed system is illustrated in Fig. \ref{fig:quark_arc}, where we can see that, our decentralized network can consist of multiple nodes that are hosted independently. A person can register as a user of a node. A user can: (a) create a channel, (b) invite other users to a channel, (c) join a channel upon invitation, (d) send messages to a channel, and (e) read messages from a channel. Each node  creates a blockchain ledger for every channel and hosts the messages of that channel in that ledger. A node hosts only the ledgers of which it's users are part of.

\vspace{-2mm}
\section{\texttt{Quarks} Protocol Flow}
\label{sec:protocol}
In this section, we present the protocol flow between different components of our proposed system. The mathematical notations and symbols used to describe this protocol flow are listed in Table \ref{table:notation}. The protocol has seven phases: i) user registration; ii) channel creation; iii) node addition; iv) channel member addition; v) channel secret key retrieval; vi) message sending and vii) receiving. 

\begin{table}[t]
\caption{Notations for Protocol Flow}
\label{table:notation}
\centering
\begin{tabular}{r|l}
\hline
\textbf{Notations}   & \textbf{Description} \\ \hline \hline
${Q_i}$              & A \texttt{Quarks} node in the network  \\ \hline
${CN_H}$             & Channel Name of Channel ${H}$   \\ \hline
${SK_H}$             & Symmetric Secret Key of  Channel ${H}$   \\ \hline
${J}_{Q_i}$          & A user registered in node ${Q_i}$ \\ \hline
${UN}_{J}$           & Username of ${J}_{Q_i}$ \\
\hline
${Adrs}_{Q_i}$       & Domain address of ${Q_i}$ \\
\hline
${Adrs}^{J}_{Q_i}$   & Domain address of ${Q_i}$ where ${J}_{Q_i}$ has registered \\
\hline
${W}_{J}$             & Wallet of ${J}_{Q_i}$ \\ \hline
${C}_{U}^{Q_i}$   & Certificate of ${J}_{Q_i}$ issued by ${Q_i}$ \\ \hline
$K_{J}$               & Public key of ${J}_{Q_i}$ \\ \hline
$K^{-1}_{J}$          & Private key of ${J}_{Q_i}$ \\ \hline
$N_i$                 & A fresh nonce   \\ \hline
${\{\}}_K$            & Encryption operation using a public key $K$  \\ \hline
${\{\}}_{K^{-1}}$     & Signature using a private key $K^{-1}$  \\ \hline
$H(.)$                & A hash function \\ \hline
${[ ]}_{\mathit{https}}$     & Communication over an HTTPS channel  \\ \hline
${msg}$ & A text message \\ \hline
${msg_{i}}$ & $ith$ $msg$ in list of all textual messages \\ \hline
${MsgL}$ & List of text messages \\ \hline
${MsgL^{K}}$ & Every element is encrypted using $K$ in this list \\ \hline
${ts}$ & A timestamp in nanoseconds \\ \hline
$\mathit{QB}$ & \texttt{Quarks} Blockchain Network \\ \hline
$\mathit{SC_{H}}$ & Smart Contract in Channel $H$ \\ \hline
$\mathit{Ldgr_{H}}$ & Ledger of Channel $H$ \\ \hline
\end{tabular}
\end{table}

\textbf{1. User Registration Phase.} [Table \ref{table:protocol_userReg}]

\indent $M_1$. Firstly, the user ($A_{Q_1}$) generates a pair of public and private keys ($K_{A}$, $K_A^{-1}$). Next, the user sends it's username, public key, and digital signature to the \texttt{Quarks} node ($Q_1$) through a \texttt{https} request. \\
\indent $M_2$. The node ($Q_1$) validates the signature, generates a digital certificate ($C_{A}^{Q_1}$) for the user, and makes an entry in the off-chain database. Finally, the node returns the certificate and the nonce ($N_1$) to the user. A fresh nonce is used in every transmission to combat replay attacks.

\begin{table}[h]
\caption{User Registration Protocol}
\label{table:protocol_userReg}
\centering
\begin{tabular}{lll}
\hline
$M_1$ & $A_{Q_1} \rightarrow Q_1:$  & ${[N_1,UN_{A}, K_{A}
, \{N_1, UN_A\}_{K_A^{-1}}]}_{https}$\\
$M_2$ & $Q_1 \rightarrow A_{Q_1}:$  & ${[N_1, C_{A}^{Q_1}]}_{https}$ \\
\hline
\end{tabular}
\end{table}

\begin{table}[h]
\caption{Channel Creation Protocol}
\label{table:protocol_createChannel}
\centering
\begin{tabular}{lll}
\hline
$M_3$ & $A_{Q_1} \rightarrow Q_1:$  & \begin{tabular}{@{}l@{}}
$[N_1, C_{A}^{Q_1},\{SK_H\}_{K_A}, $\\
$CN_H, \{N_1, CN_H\}_{K_A^{-1}}]_{https}$
\end{tabular}\\
$M_4$ & $Q_1 \rightarrow A_{Q_1}:$  & $ {[N_1]}_{https}$ \\
\hline
\end{tabular}
\end{table}

\textbf{2. Channel Creation Phase.} [Table \ref{table:protocol_createChannel}] \\
\indent $M_3$. The user ($A_{Q_1}$) requests the node ($Q_1$) to create a new channel through a \texttt{https} request. The request includes the name of the new channel ($CN_H$), certificate ($C_{A}^{Q_1}$), signature, and the channel secret key encrypted with the user's public key ($\{SK_H\}_{K_A}$). The secret key will be generated by the user creating the channel.\\
\indent $M_4$. The node ($Q_1$) validates the user's certificate ($C_{A}^{Q_1}$), digital signature, and then creates a new ledger for this channel. Furthermore, the node sets up the smart contract for this channel and initiates the ledger for messaging. At initiation, the smart contract stores the encrypted secret key ($SK_H$) for this channel.  \vspace{-3mm} \\

\textbf{3. Node Addition Phase.} [Table \ref{table:protocol_addNode} and Algorithm-\ref{algo:smartContract} (lines 8-14)] \\
\indent $M_5$. A member ($A_{Q_1}$) of the channel sends a request to the node ($Q_1$) for federating with a new node ($Q_2$) for the channel. The request includes member's certificate, signature, name of the channel, and the domain address of the new node. \\
\indent $M_6$. The node ($Q_1$) validates the request, fetches the certificate, and invokes smart contract function for adding the node ($Q_2$) in the channel. The node ($Q_1$) then sends it's own certificate ($C_{A}^{Q_1}$) and the new node's certificate ($C_{A}^{Q_2}$). \\
\indent $M_7$. The smart contract function validates the request and add the new node's certificate (Algorithm \ref{algo:smartContract}, lines 9-10) in the authorized list of nodes and sends back a response (${N_2}$) to the node ($Q_1$). \\
\indent $M_8$. After receiving the success response from the smart contract function, the node asks the new node to join the channel, synchronizes the ledger ($Ldgr_H$), and the smart contract ($SC_H$). The message contains the encrypted ledger and the smart contract. \\
\indent $M_9$. The new node ($Q_2$) validates the request, creates a replica of the ledger, and sets up the smart contract in the ledger. Finally, the new node starts synchronizing the ledger and the smart contract with all the participating nodes of the channel and sends back a nonce (${N_3}$) to the node($Q_1$). \\ 
\indent $M_{10}$. Upon successful addition of a new node ($Q_2$) in the channel, the node ($Q_1$) sends back a ``success message" to the the requesting member ($A_{Q_1}$) of the channel. \vspace{-2mm} \\

\begin{table}[t]
\caption{Node Addition Protocol}
\label{table:protocol_addNode}
\centering
\begin{tabular}{lll}
\hline
$M_5$ & $A_{Q_1} \rightarrow Q_1:$  & \begin{tabular}{@{}l@{}}
${[N_1, C_{A}^{Q_1},}$\\
${Adrs}_{Q_2}, CN_H, \{N_1, CN_H\}_{K_A^{-1}}]_{https}$
\end{tabular}\\

$M_6$ & $Q_1 \rightarrow SC_H:$  & \begin{tabular}{@{}l@{}}
${[N_2, C_{Q_1}, C_{Q_2}]}_{https}$
\end{tabular}\\

$M_7$ & $SC_H \rightarrow Q_1:$  & \begin{tabular}{@{}l@{}}
${[N_2]}_{https}$
\end{tabular}\\

$M_8$ & $Q_1 \rightarrow Q_2:$  & \begin{tabular}{@{}l@{}}
${[N_3, C_{Q_1}, Ldgr_H, SC_H ]}_{https}$
\end{tabular}\\

$M_9$ & $Q_2 \rightarrow Q_1:$  & ${[N_3]}_{https}$ \\

$M_{10}$ & $Q_1 \rightarrow A_{Q_1}:$  & ${[N_1, \text{success msg}]}_{https}$ \\

\hline
\end{tabular}
\end{table}

\IncMargin{1em}
\begin{algorithm}
\SetAlgoLined
\caption{Smart Contract}
\label{algo:smartContract}
\small
\textbf{Input:} $req$ \Comment{function name and parameters}\\
\textbf{Output:} $resp$ \Comment{output of function}\\
\SetKwBlock{Begin}{}{}
\Begin(\textbf{Start})
{
    $ChNodes \leftarrow \texttt{GetChNodeCerts()}$\\
    \Comment{get certificate list of the nodes of the channel}\\
    
    $ChUsers \leftarrow \texttt{GetChUsersCerts()}$\\
    \Comment{get certificate list of the users of the channel}\\
    
    \SetKwProg{addNode}{\textit{fn} addNode}{($ N_2, C_{Q_1}, C_{Q_2}$)}{}
    \addNode{}{
        
        \uIf{$C_{Q_1} \in ChNodes$}
        {
        $ChNodes \leftarrow ChNodes \cup C_{Q2}$\\
        \Comment{add the new certificate to the set}\\
        
        \texttt{PutChNodeCerts($ChNodes$)}\\
        \Comment{put the updated list in the ledger}
        
        \KwRet $N_2$
        }
    }
    
    \SetKwProg{addMember}{\textit{fn} addMember}{($N_3,C_{Q_1}, C_{A}^{Q_1}, C_{B}^{Q_1}, \{SK_H\}_{K_B}$)}{}
    
    \addMember{}{
          \uIf{$C_{Q_1} \in ChNodes \land C_{A}^{Q_1} \in ChUsers$}
            {
            \texttt{PutSK($C_{B}^{Q_1}, \{SK_H\}_{K_B}$)}\\
            \Comment{put encrypted secret key in the ledger}\\
            
            \KwRet $N_3$
            }
    }
    
    \SetKwProg{getChSK}{\textit{fn} getChannelSK}{($ N_2, C_{Q_1}, C_{A}^{Q_1}$)}{}
    \getChSK{}{
        \uIf{$C_{Q_1} \in ChNodes \land C_{A}^{Q_1} \in ChUsers$}
        {
        $\{SK_H\}_{K_A} \leftarrow$ \texttt{GetSK($C_{A}^{Q_1}$)}\\
        \Comment{get encrypted secret key from ledger}\\
        
        \KwRet $N_2, \{SK_H\}_{K_A}$
        }
    }
    
    \SetKwProg{sendMsg}{\textit{fn} sendMsg}{($ N_2, C_{Q_1}, C_{A}^{Q_1}, \{msg\}_{SK_H}$)}{}
    \sendMsg{}{
        \uIf{$C_{Q_1} \in ChNodes \land C_{A}^{Q_1} \in ChUsers$}
        {
        $\mathit{ts} \leftarrow \texttt{GetTs()}$\\
        \Comment{get current timestamp in nanoseconds}
        
        \texttt{PutState($ts, \{msg\}_{SK_H}$)}\\
        \Comment{store encrypted message in the ledger}\\
        
        \KwRet $\mathit{N_2}$
        }
    }
    
    \SetKwProg{readMsg}{\textit{fn} readMsg}{($ N_2, C_{Q_1}, C_{A}^{Q_1}, ts$)}{}
    \readMsg{}{
          \uIf{$C_{Q_1} \in ChNodes \land C_{A}^{Q_1} \in ChUsers$}
        { $\mathit{ts_f} \leftarrow ts$\\
        
        $\mathit{ts_t} \leftarrow \texttt{GetTs()}$\\
        \Comment{get current timestamp in nanoseconds}
        
        $MsgL^{SK_H} \leftarrow$ \texttt{GetStateByRange($ts_f, ts_t$})\\
        \Comment{get messages sent between $ts_f$ and $ts_t$}\\
        
        \KwRet $\mathit{N_2, MsgL^{SK_H}}$
        }
    }

}
\end{algorithm}

\textbf{4. Channel Member Addition Phase.} [Table \ref{table:protocol_addMember} and Algorithm-\ref{algo:smartContract} (lines 15-19)]

\indent $M_{11}$. At first, a member ($A_{Q_1}$) of the channel sends a request to the node ($Q_1$) to add a user ($UN_B$) to the channel. The request contains certificate of the member ($C_{A}^{Q_1}$), the signature of the member, new username ($UN_B$), and the node's address (${{Adrs}^{B}_{Q_1}}$) of the user. \\
\indent $M_{12}$. After request validation, the node fetches the user's certificate using user's node address and username. If the new member was registered on the same node, the node will be able to fetch it from it's local database. Otherwise, the node has to request the user's node to share the certificate of the user. Next, the node sends the requesting member ($A_{Q_1}$) the public key ($K_B$) of the user.

\begin{table}[t]
\caption{Add Member Protocol}
\label{table:protocol_addMember}
\centering
\begin{tabular}{lll}
\hline
$M_{11}$ & $A_{Q_1} \rightarrow Q_1:$  & \begin{tabular}{@{}l@{}}
$ {[N_1, C_{A}^{Q_1}, UN_B, {Adrs}^{B}_{Q_1}}$\\
$CN_H, \{N_1, CN_H\}_{K_A^{-1}}]_{https}$
\end{tabular}\\
$M_{12}$ & $Q_1 \rightarrow A_{Q_1}:$  & ${[N_1, K_B]}_{https}$ \\
$M_{13}$ & $A_{Q_1} \rightarrow Q_1:$  & ${[N_2,  \{SK_H\}_{K_B}}]_{https}$ \\
$M_{14}$ & $Q_1 \rightarrow SC_H:$  & \begin{tabular}{@{}l@{}}
${[N_3,C_{Q_1}, C_{A}^{Q_1},}$\\
$C_{B}^{Q_1},  \{SK_H\}_{K_B}]_{https}$
\end{tabular}\\
$M_{15}$ & $SC_H \rightarrow Q_1:$  & ${[N_3]}_{https}$ \\
$M_{16}$ & $Q_1 \rightarrow A_{Q_1}:$  & ${[N_2, \text{success msg}]}_{https}$ \\
\hline
\end{tabular}
\vspace{-0.3cm}
\end{table}

\indent $M_{13}$. The member ($A_{Q_1}$) now encrypts the channel's secret key ($SK_H$) using the received public key ($K_B$) and send it to the node ($Q_1$). \\
\indent $M_{14}$. The node invokes the smart contract of the channel to store the encrypted channel secret key. The node then sends the certificate of itself, the requesting member's certificate, the new user's certificate, and the encrypted channel secret key. \\
\indent $M_{15}$. The smart contract function (Algorithm \ref{algo:smartContract}, line 17) stores the encrypted secret key in the ledger. As the key was encrypted using user's public key, only the entity having the private key will be able to decrypt this secret key. \\
\indent $M_{16}$. Finally, after successfully invoking add member function of the smart contract, the node sends the requesting member a success message. \vspace{-2mm} \\

\textbf{5. Channel Secret Key Retrieval Phase.} [Table \ref{table:protocol_getChannelSK} and Algorithm-\ref{algo:smartContract} (lines 20-24)] \\
\indent $M_{17}$. The user requests the node to get the encrypted channel secret key. The request includes name of the channel, user's certificate, and the digital signature. \\
\indent $M_{18}$. The node validates the user's request and checks if the channel exists. If the channel is found, the node invokes smart contract function (Algorithm \ref{algo:smartContract}, lines 21-22) for retrieving the encrypted secret key ($SK_H$) of the channel.  \\
\indent $M_{19}$. Upon validating the parameters, smart contract retrieves the encrypted secret key from the ledger and sends back the encrypted key to the node. \\
\indent $M_{20}$. The node ($Q_1$) responds to user's request by sending back the encrypted key. The user ($A_{Q_1}$) will be able to decipher the key ($SK_H$) using it's private key ($K^{-1}$). \vspace{-0.3cm}\\

\begin{table}[h]
\caption{Get Channel Secret Key Protocol}
\label{table:protocol_getChannelSK}
\centering
\begin{tabular}{lll}
\hline
$M_{17}$ & $A_{Q_1} \rightarrow Q_1:$  & \begin{tabular}{@{}l@{}}
${[N_1, C_{A}^{Q_1},}$ \\
$CN_H, \{N_1, CN_H\}_{K_A^{-1}}]_{https}$
\end{tabular}\\
$M_{18}$ & $Q_1 \rightarrow SC_{H}:$  & $ {[N_2,C_{Q_1}, 
C_{A}^{Q_1}}]_{https}$ \\
$M_{19}$ & $\mathit{SC}_{H} \rightarrow Q_1:$  & ${[N_2, \{SK_H\}_{K_A}]}_{https}$ \\
$M_{20}$ & $Q_1 \rightarrow A_{Q_1}:$  & $ {[N_1, \{SK_H\}_{K_A}]}_{https}$\\
\hline
\end{tabular}
\vspace{-3mm} 
\end{table}

\textbf{6. Message Sending Phase.} [Table \ref{table:protocol_sendMessage} and Algorithm-\ref{algo:smartContract} (lines 25-31)] \\
\indent $M_{21}$. The user ($A_{Q_1}$) sends, the message encrypted with the channel secret key, user's certificate, name of the channel, and the signature to the node ($Q_1$).  \\
\indent $M_{22}$. The node ($Q_1$) validates the request, invokes the smart contract function (Algorithm \ref{algo:smartContract}, line 25) for sending messages in the channel.  \\
\indent $M_{23}$. The smart contract ($SC_H$) function validates the user's membership in the channel (Algorithm \ref{algo:smartContract}, line 26) and if succeeds, puts the encrypted message in the ledger. \\
\indent $M_{24}$. Finally, the node ($Q_1$) sends back a success message to the user ($A_{Q_1}$) after a successful smart contract invocation.
\begin{table}[!h]
\caption{Send message protocol}
\label{table:protocol_sendMessage}
\centering
\begin{tabular}{lll}
\hline
$M_{21}$ & $A_{Q_1} \rightarrow Q_1:$  & \begin{tabular}{@{}l@{}}
${[N_1, C_{A}^{Q_1}, \{msg\}_{SK_H}}$\\
$CN_H, \{N_1, CN_H\}_{K_A^{-1}
}]_{https}$
\end{tabular}\\
$M_{21}$ & $Q_1 \rightarrow \mathit{SC}_{H}:$  &
\begin{tabular}{@{}l@{}}
$[N_2, C_{Q_1}, C_{A}^{Q_1},$\\
$\{msg\}_{SK_H}]_{https}$
\end{tabular}\\
$M_{23}$ & $\mathit{SC}_{H} \rightarrow Q_1:$  & ${[N_2]}_{https}$ \\
$M_{24}$ & $Q_1 \rightarrow A_{Q_1}:$  & ${[N_1, \text{success msg}]}_{https}$\\
\hline
\end{tabular}
\end{table}

\textbf{7. Message Reading Phase.} [Table \ref{table:protocol_readMessage} and Algorithm-\ref{algo:smartContract} (lines 32-39)] \\
\indent $M_{25}$. The user ($A_{Q_1}$) asks the node ($Q_1$) to fetch all messages in a period of time. The
request includes the channel's name, certificate, signature, and a timestamp (in nanoseconds). \\
\indent $M_{26}$. After validating the request, the node invokes the smart contract function (Algorithm \ref{algo:smartContract}, lines 32) to read the contents of the channel. The node passes the timestamp along with the certificates of the node and the user as function parameters. \vspace{0.5mm} \\
\indent $M_{27}$. Once the validation of parameters is done, the smart contract queries the encrypted messages from the ledger and sends back the message list ($MsgL^{SK_H}$) to the node. \vspace{0.5mm} \\
\indent $M_{28}$. The node ($Q_1$) sends back the message list to the user ($A_{Q_1}$). Finally, the user will be able to read the message contents by decrypting the messages using channel secret key. 

\begin{table}[h]
\caption{Read message protocol}
\label{table:protocol_readMessage}
\centering
\begin{tabular}{lll}
\hline
$M_{25}$ & $A_{Q_1} \rightarrow Q_1:$  & \begin{tabular}{@{}l@{}}
${[N_1, C_{A}^{Q_1}, ts}$\\
$CN_H, \{N_1, CN_H\}_{K_A^{-1}
}]_{https}$
\end{tabular}\\
$M_{26}$ & $Q_1 \rightarrow \mathit{SC}_{H}:$  &
\begin{tabular}{@{}l@{}}
$[N_2, C_{Q_1}, C_{A}^{Q_1},$\\
$ts]_{https}$
\end{tabular}\\
$M_{27}$ & $\mathit{SC}_{H} \rightarrow Q_1:$  & ${[N_2, MsgL^{SK_H}]}_{https}$ \\
$M_{28}$ & $Q_1 \rightarrow A_{Q_1}:$  & ${[N_1,MsgL^{SK_H}
]}_{https}$\\
\hline
\end{tabular}
\end{table}

\begin{figure}[ht]
\centering
\includegraphics[width=2.8in, height=1.2in]{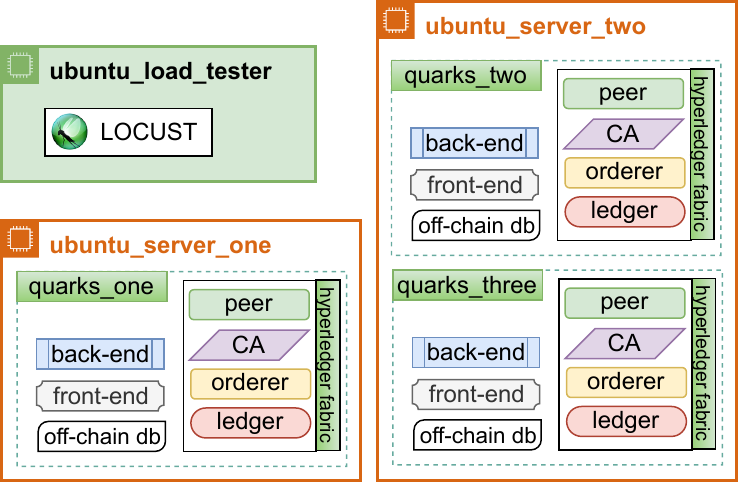}
\caption{Proof-of-Concept of \texttt{Quarks} System}
\label{fig:quark_chat_poc}
\end{figure}


\section{Implementation}
\label{sec:implementation}
We have implemented a proof-of-concept and analyzed it's performance. We have recorded the throughput and latency of (a) Send Message and (b) Read Message; simulating with increasing loads using an open source load testing tool named Locust \cite{locust}.
\begin{figure*}[t]
\centering
\includegraphics[width=7.2in]{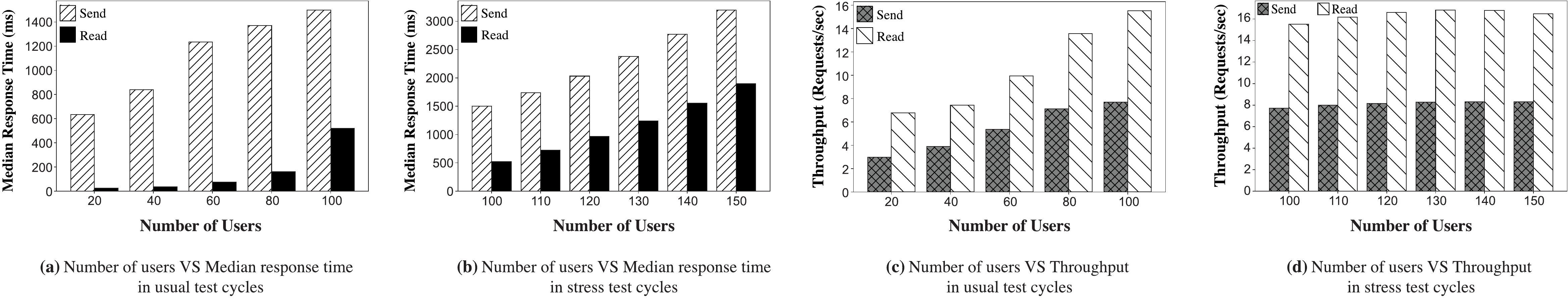}
\caption{Performance evaluation of Quarks (Median Response Time and Throughput)}
\label{fig:performance_quarks}
\end{figure*}


\subsection{Proof of Concept (PoC)}

We leveraged \textit{Hyperledger Fabric} \cite{HyperledgerFabric} to implement the blockchain network. Here, every \texttt{Quarks} node have their own Certificate Authority (CA), an Orderer, and multiple peers. We developed fabric chaincode using golang \cite{golang} for implementing \textit{Smart Contract}. For \textit{Ledger}, a fabric-channel \cite{fabricChannel} is opened for every quarks channel. We built a backend service using NodeJs \cite{nodejs} so that our clients can interact with the blockchain. We used ReactJS \cite{reactjs} for our \textit{Frontend} and MongoDB \cite{mongodb} for the off-chain database. \\

\vspace{-0.1cm}
\subsection{Performance Analysis}

For testing the performance of our PoC, we have set up a controlled simulation environment targeting a maximum of 100 users. Figure \ref{fig:quark_chat_poc} shows the structure of our environment. It contains a Locust and 2 \texttt{Ubuntu} servers containing 3 nodes of the \texttt{Quarks} system. The simulation is carried out in a series of cycles that vary depending on the number of users. First, we start with 20 users and add 20 more users in each cycle till 100 users. Each cycle contains the following three steps: i) each simulated user sends and reads messages from the \texttt{Quarks} using a REST API; ii) for each API request, the response time is measured (in ms); and iii) on the Locust end, the throughput is measured by calculating the number of requests served per second. 

After simulating for 100 users, we also performed a stress test on our environment. We added 50 additional users in 5 more cycles to check if the system can handle excessive loads. Figure \ref{fig:performance_quarks}(a) and \ref{fig:performance_quarks}(b) shows the median response time for all our test cycles. For usual test cycles (20-100 users), the response time is decent considering the decentralized architecture. It increases linearly as the number of users increases. And for stress testing cycles, the response time is respectively higher. However, our system can handle all requests of up to 150 users with its limited capacity. In figure \ref{fig:performance_quarks}(c), the system's throughput increases with the number of users in usual test cycles. As the number of users grows, so does the number of requests per second (throughput). However, in stress test cycles figure \ref{fig:performance_quarks}(d), the throughput reaches a saturated level. This indicates that our PoC has reached its maximum capacity. 

Though we tested our PoC in a small simulation environment, it showed 100\% availability and decent performance in sending and reading messages. We believe that with proper resources, \texttt{Quarks} could be scaled to use as a decentralized messaging system. Also, in the upcoming era of 5G and the increasing popularity of DApps (Decentralized Applications), \texttt{Quarks} will be more feasible to use and can be integrated with other decentralized systems for secured message sharing.  

\subsection{Informal Security analysis of \texttt{Quarks}}
\label{security_analysis}
\textbf{Confidentiality.} Every message in the ledger in \texttt{Quarks} is encrypted with a secret key unique to a channel. This channel's secret key is only known to the members of the channel. This guarantees protection against potential data breaches. \\
\textbf{Integrity.} In \texttt{Quarks} all the messages are stored on-chain. That means it is impossible to delete or tamper the previous messages. \\
\textbf{Non-Repudiation.} As every message-write is digitally signed by the members and signatures are being verified by smart contract, Quarks diminishes repudiation attempts. \\
\textbf{Authentication and Authorization.} Users interact with the \texttt{Quarks} by authenticating with their private keys which enables the \texttt{Quarks} nodes to prevent unauthorized access to the users’ data. \\
\textbf{Resilient to DDoS Attacks.} DDoS attackers may successfully make a single node unavailable for some time; however, the decentralized nature of the \texttt{Quarks} network will keep the service intact.

\section{Conclusion}
\label{sec:conclusion}
To ensure private and secure communication over a decentralized network, we have designed and developed a messaging system leveraging distributed ledger technology. Going forward, we believe that our design will empower individuals to set-up nodes and communicate with their peer nodes securely. Our implemented PoC assured us of the feasibility of such systems. We envision that, our novel approach to solve the current security issues of the existing messaging systems will open up new school of thoughts and pioneer the next generation messaging services that we would be using for secure communication and collaboration.

\bibliographystyle{IEEEtran}
\bibliography{IEEEabrv,references}

\end{document}